\documentclass[a4paper,11pt]{article}
\usepackage{jheppub}
\pdfoutput=1
\usepackage{enumitem}
\usepackage[activate=true]{microtype}
\usepackage{soul,color}
\usepackage{ifthen}
\usepackage{booktabs}
\usepackage{xspace}

\usepackage[T1]{fontenc}
\usepackage[utf8]{inputenc}

\newboolean{@draft}
\setboolean{@draft}{false}
\ifthenelse{\boolean{@draft}}{  }{  }

\ifthenelse{\boolean{@draft}}{
\usepackage{draftwatermark}
\SetWatermarkText{DRAFT}
\SetWatermarkScale{3.5}
}{}

\newcommand{\abinv}{\mbox{\ensuremath{\,\text{ab}^{-1}}}\@\xspace}
\newcommand{\ie}{i.e.\@\xspace}

\newcommand{\cf}{cf.\xspace}

\newcommand{\km}{\ensuremath{\,\text{km}}\xspace}

\newcommand{\TeV}{\ensuremath{\,\text{Te\hspace{-.08em}V}}\xspace}
\newcommand{\GeV}{\ensuremath{\,\text{Ge\hspace{-.08em}V}}\xspace}

\title{Brazilian input to the European Strategy for Particle Physics Update}

\author[a]{(Eds.) U. de Freitas Carneiro da Graça}
\author[b]{G. Gil da Silveira}
\author[c]{C. Jahnke}
\author[d]{A. Lessa}
\author[e]{H. Malbouisson}
\author[f]{E. E. Purcino de Souza}
\author[g]{M. S. Rangel}
\author[h]{T. R. Fernandez Perez Tomei}
\author[a]{A. Vilela Pereira}

\contributor[i]{Contributors: J. Citadini}
\contributor[i]{M. F. A. Brito}
\contributor[e]{S. Fonseca De Souza}
\contributor[a]{C. Hensel}
\contributor[a]{E. Silva Junior}

\contact[g]{Contact: L. de Paula}

\emailAdd{leandro.de.paula@cern.ch}

\affiliation[a]{
Centro Brasileiro de Pesquisas Físicas~(CBPF)
}
\affiliation[b]{
Universidade Federal do Rio Grande do Sul~(UFRGS)
}
\affiliation[c]{
Universidade Estadual de Campinas~(UNICAMP)
}
\affiliation[d]{
Universidade Federal do ABC~(UFABC)
}
\affiliation[e]{
Universidade do Estado do Rio de Janeiro~(UERJ)
}
\affiliation[f]{
Universidade Federal da Bahia~(UFBA)
}
\affiliation[g]{
Universidade Federal do Rio de Janeiro~(UFRJ)
}
\affiliation[h]{
Universidade Estadual Paulista~(Unesp)
}
\affiliation[i]{
Centro Nacional de Pesquisa em Energia e Materiais~(CNPEM) 
}

\abstract{
 The Brazilian High-Energy Physics (HEP) community has expanded remarkably since its first involvement at CERN and Fermilab in the 1980s. Its recent organization under the Brazilian Network for High-Energy Physics (RENAFAE), since 2008, has further strengthened its scientific and technological goals, particularly in detector instrumentation, computing, and industry partnerships. In 2024, Brazil became an Associate Member State of CERN, opening new opportunities for deeper engagement in accelerator and detector R\&D. This input to the 2026 update of the European Strategy for Particle Physics highlights Brazil's current participation in LHC experiments as well as ongoing developments in detector and accelerator technology, and details the community's view towards future colliders. The potential for expanded scientific and industrial collaborations between Brazil and CERN is also discussed.
}

\begin{document}

\maketitle

\section{Introduction}
\ifthenelse{\boolean{@draft}}{ \textcolor{red}{responsável:  Thiago , Antonio}\\ }

The Brazilian High-Energy Physics community has grown steadfastly in the last few decades. 
In the context of modern collider experiments,
Brazilian physicists have collaborated with CERN since the 1980s, starting from the SPS NA22 experiment, and the DELPHI experiment at the Large Electron-Positron Collider~(LEP) from 1990;
Brazilian institutes have also participated, since 1985, in the Tevatron's D\O{} experiment and previously in Fermilab's fixed-target experiments.
Since 2008, the community has been organized under the umbrella of the \textit{Brazilian Network for High-Energy Physics}~(RENAFAE)~\cite{renafae}, established with the goal of promoting HEP research in Brazil, and the technological development for and from HEP in partnership with the local industry. RENAFAE aims to coordinate HEP activities fostering common actions with funding agencies, including accelerator-based as well as neutrino and astroparticle physics experiments.

In 2022, the \textit{National Council for Scientific and Technological Development}~(CNPq) approved for the first time a five-year plan for CERN-related activities in the form of a \textit{National Institute for Science and Technology}~(INCT). INCT CERN-BR~\cite{inct} has been set up with an executive structure formed by several working groups ranging from instrumentation in detectors and electronics, computing infrastructure, training and outreach.

While the INCT CERN-BR budget is limited ($\sim 1\,\textrm{MCHF}$), it has strengthened the interaction within the community and has launched new initiatives, targeting larger-scale funding in cooperation with the industry. The aforementioned initiatives are critical to support the local HEP community \& industry in the dawn of Brazil’s association with CERN, formalized in March 2024. 

\section{Brazilian landscape in detectors and accelerator R\&D} \ifthenelse{\boolean{@draft}}{ \textcolor{red}{responsável: ... , Antonio}\\
\textcolor{red}{contribuição: James / Marisa - CNPEM}\\ }

\subsection{Detector R\&D}
\label{sec:detectorRD}

Although the INCT CERN-BR initiative is relatively recent, a significant increase in cooperation among researchers in detector R\&D has already been observed. This collaboration covers the development of gaseous detectors, including large-area detectors, micropattern gaseous detectors, eco-friendly gas mixtures and gas regeneration; semiconductor sensors, especially ultra-fast silicon detectors, artificial diamond sensors, characterization techniques and applications, e.g. for X-ray detection; photon detectors, in particular for large-area tracking; sensor read-out \& front-end electronics; high-speed electronics for online systems.
Brazilian institutes are members of Detector R\&D~(DRD) collaborations at CERN.
Expansion of the Brazilian capacity for prototyping and assembly of detectors is foreseen.

\subsection{Accelerator R\&D}
\label{sec:acceleratorRD}

Accelerator research and related applications are spearheaded by the \textit{Brazilian Center for Research in Energy and Materials}~(CNPEM), home to the 4$^{th}$ generation synchrotron light source \textit{SIRIUS}~\cite{sirius}.

CNPEM has been investing in the creation of a hadron accelerator, with the present goal of building a proton accelerator capable of producing radioisotopes and being used in proton therapy at a final energy of around 250 MeV. In this context, there is a desire to support and acquire competence in radiofrequency systems (amplifiers, electronics, and cavities), magnetic systems (with a focus on superconductors), and electronics for beam monitoring systems and detectors. 

CNPEM remains interested in partnering with CERN to expand its capabilities and with direct involvement in its programs. In this context, we highlight the ongoing deployment of engineering team members to work directly on projects related to vacuum systems, infrastructure, and materials for upgrades at the Large Hadron Collider~(LHC), as well as vacuum chamber modeling and new RF cavities for the FCC-ee project. 

Additionally, efforts are underway for the development of the superconducting magnet for ALICE 3, including support for the establishment of a Brazilian supplier for the Rutherford-type Nb-Ti/Al wire.

Beyond the ongoing activities, CNPEM reiterates its interest in CERN's consultancy -- as has previously been done for SIRIUS' Superconducting Wavelength Shifter insertion device and the proton linear accelerator's quadrupole RF cavity -- in addition to new joint technological developments in the areas of accelerator physics, optics, and lattice design; etching and cleaning process for materials deposition; vacuum chambers (and joints techniques); NEG coating (for vacuum applications); niobium deposition (as well as NbTi, Nb3Sn) for RF cavities; front-end electronics, ASIC design and testing; magnet simulation and manufacturing (superconducting magnets); magnetic measurements; cryogenics; technology for Superconducting Rutherford cable; beam monitors; superconducting power transfer lines; power supplies; radiation monitors; alignment; high power lasers.


\section{Collider project at CERN}

\subsection{Current interests of Brazilian groups}
\ifthenelse{\boolean{@draft}}{ \textcolor{red}{responsável: Cristiane, Gustavo}\\ }

Brazil has been actively involved in LHC experiments and their upgrades, participating in all four major collaborations: ALICE, ATLAS, CMS, and LHCb.
Below, we highlight the main physics interests of the participants in each collaboration.

The ALICE groups in Brazil study the Quark-Gluon Plasma (QGP) through heavy-ion collisions (HIC), investigating global properties of this medium through the measurement of strangeness production and microscopic properties through hard probes, as heavy quark production and jet quenching. The Brazilian groups have been applying Machine Learning (ML) techniques to improve the reconstruction of strange baryons, quarkonia identification, and jet quenching, enhancing the precision in ALICE.  These groups are also involved in developing scientific instrumentation for detector upgrades. An important milestone was the development of the SAMPA ASIC, the new readout electronics of the ALICE Time Projection Chamber (TPC) and the Muon Chamber, fully designed in Brazil. The ALICE TPC upgrade for the LHC Run 3 replaced the multi-wire proportional chambers with Gas Electron Multiplier (GEM) technology, allowing for the continuous and higher rate provided by the SAMPA chip. Coupled with this development, the Brazilian groups are proposing new approaches and applications for GEM detectors as well. Brazilian institutions are also involved in developing the Forward Calorimeter (FoCal) aimed for the LHC Run 4 period. FoCal will provide high-granularity electromagnetic calorimetry in the forward region, crucial for studying small-x QCD physics and gluon saturation effects. In the future, for LHC Run 5, the ALICE detector will be completely replaced by ALICE 3, a full-silicon detector with improved acceptance and vertex resolution. ALICE 3 will enhance QGP studies with high-precision heavy-quark and quarkonia measurements and the ALICE Brazilian groups are planning to contribute to the Time-Of-Flight (ToF) system that uses ultra-fast silicon detectors. Simulations of the performance of the ToF system for particle identification have already started.


The ATLAS group in Brazil contributes to multiple aspects of the experiment, from hardware development to physics analysis. In physics analysis, Brazilian researchers focus on the search for non-resonant and resonant Higgs boson pair production, 
precision measurements of the Standard Model~(SM), with an emphasis on tau final states, heavy-flavor tagging, and searches for beyond Standard Model~(BSM) physics, including axion-like particle (ALP) searches.
In scientific instrumentation, Brazilian institutes have made significant contributions to the electromagnetic liquid argon calorimeter, particularly in developing the online trigger system. Extensive work has been carried out on electron/photon triggers, mainly in the design, development and deployment of reconstruction, identification, data quality and monitoring algorithms for Run 3, also involving ML applications.
Prototypes for the High Luminosity LHC~(HL-LHC) have also been developed using ML for high-speed embedded electronics, which are dedicated to online event identification.
Furthermore, the Brazilian group participates in the upgrade of the ATLAS detector for HL-LHC.
Brazilian institutes are actively engaged in the R\&D, characterization and qualification of Low Gain Avalanche Diode~(LGAD) silicon sensors, which are used in the High Granularity Timing Detector~(HGTD) to be installed in the ATLAS experiment.
These detectors will improve time resolution in tracking, mitigating pile-up effects at high luminosities.
In a related application, LGAD sensors are also being developed for time-resolved synchrotron light detection in collaboration with CNPEM.
Beyond hardware and data analysis, Brazilian researchers are also involved in offline computing, including ML applications for event reconstruction and real-time data filtering.

The Brazilian groups in the CMS experiment are engaged in instrumentation and physics analyses, with participation in the Resistive Plate Chambers~(RPC) of the Muon System, the Electromagnetic Calorimeter~(ECAL) and the Precision Proton Spectrometer~(PPS). Contributions on assembly, installation, testing and monitoring were made for the various detector upgrades. Activities in local laboratories include the R\&D of solid-state timing sensors as well as gas mixture and signal formation studies in gaseous detectors. The CMS groups are responsible for OpenIPMC, a free and open-source IPMI (Intelligent Platform Management Interface) controller: a complete package designed for Advanced Telecommunications Computing Architecture~(ATCA) boards. Approximately 1,100 IPMC cards will be needed by the experiment.
The broad physics program covers several topics in proton-proton and HIC to deepen our understanding of fundamental interactions and probe for possible signs of new physics. Topics include searches for heavy-mass mediators as well as signals compatible with exotic long-lived particles. Such searches often rely on cutting-edge data analysis strategies, including the implementation of ML techniques to enhance signal sensitivity and background discrimination. Searches of BSM signals that manifest through anomalous gauge couplings and rare electroweak processes are also carried out. Feasibility studies are performed for the conditions at the HL-LHC. Jet physics is another area of intense research, with the development of novel techniques for analyzing jet substructure and properties, as is heavy flavour physics with measurements of $b$ and $c$ hadrons production rates and properties.
In the context of HIC, measurements of photon-induced processes and the dynamics of the QGP are performed, offering key insights into the thermodynamic properties of strongly interacting matter, such as the speed of sound, as well as indirect access to its transport properties through the study of flow and correlations among the final-state particles produced in the collisions.

Members of the LHCb group in Brazil have made significant contributions to both the physics programme and technological development for the experiment. In data analysis, the group plays a key role in studying Charge-Parity (CP) violation in the decays of B and D mesons, particularly in final states involving three light charged hadrons. Their research goals include the study of rare decays, branching fraction measurements, and searches for violations of CP symmetry and lepton flavor universality. The group is also pioneering analyses within the experiment in areas such as jet measurements and diffractive physics, including the observation of exotic mesons and searches for new physics.
In technological development for the experiment, the Brazilian team is actively engaged in three major projects: the Vertex Locator (VELO), the Scintillating Fibre Tracker (SciFi), and Real-Time Analysis (RTA). The group played a significant role in the first major LHCb upgrade (Upgrade I) and is now preparing for the upcoming Upgrade II. For the VELO subdetector, their contributions include developing and testing of silicon pixel sensors, as well as improving the detector's mechanical and thermal stability. In the SciFi project, the group has been instrumental in all stages of the readout electronics development, from the design of the boards to their commissioning. For RTA, the group is responsible for jet reconstruction in the second level of the trigger system and photon reconstruction in the first level of the trigger, which is implemented on graphical processing units (GPUs). These efforts underscore Brazil’s strong presence in the LHCb experiment, contributing to both technological advancements and fundamental physics results.

\subsection{Review of future collider options}
\label{sec:reviewColliderOptions}
\ifthenelse{\boolean{@draft}}{ \textcolor{red}{responsável: Helena (LC) + Andre Lessa (FCC) }\\ }

\subsubsection*{Future Circular Collider}

The Future Circular Collider~(FCC) is one of the leading proposals for the next generation of colliders at CERN~\cite{Bernardi:2022hny}.
Its main infrastructure requires building a $\sim 90\km$ underground tunnel located close to Geneva, Switzerland.
The tunnel would first host a high-luminosity lepton collider (FCC-ee), operating for over a decade at several center-of-mass energies.
At a later stage, a high-energy hadron collider (FCC-hh) would operate at $85\textrm{--}100\,\TeV$ and collect a total luminosity of $\sim 20\,\text{ab}^{-1}$ over 25 years of operation.

The proposal envisages an FCC-ee operation in distinct stages. It would first run at $91\GeV$ to enable high-precision studies of the Z boson, followed by a run at $160\GeV$ to study the W boson. A further upgrade would allow it to reach $240\GeV$, focusing on Higgs measurements, exploiting the $HZ$ production channel.
Additional upgrades would eventually allow center-of-mass energies of around $350\GeV$, ideal for studying top physics.
The runs at center-of-mass energies close to the $Z$ and $W$ poles will reduce uncertainties of electroweak precision observables by more than an order of magnitude~\cite{Blondel:2021ema}. For higher center-of-mass energies, the program will allow increasing the precision of measurements of the Higgs couplings to the subpercent level.
Therefore, the FCC-ee will push the high precision frontier, setting strong constraints on new physics contributions to SM observables.

Following the FCC-ee, the FCC-hh would be constructed in the same tunnel as a $\mathcal{O}(100\TeV)$ proton-proton collider to pursue the energy frontier further. While the FCC-ee is a high-precision collider, the FCC-hh aims to be a discovery machine.
It has the potential to significantly improve measurements of the Higgs self-coupling, with an expected precision between $\sim 3\textrm{--}8\%$~\cite{Mangano:2020sao}, and provide the best possible precision for rare Higgs decays. 
The high collision energy would enable the discovery of new particles at the multi-TeV scale and allow a detailed exploration of any new physics signals which could have started appearing at the FCC-ee.

\subsubsection*{Linear Collider Facility at CERN}
The Linear Collider Facility at CERN~(LCF@CERN)~\cite{LCF:whitepaper} is a proposal for an electron-positron linear collider to measure the Higgs boson properties with high precision.

In the Brazilian scientific community’s vision, this could be presented as an alternative in an eventual impossibility of construction of the FCC-ee.
The LCF@CERN is an effort from different organizations such as the International Linear Collider~(ILC), the Compact Linear Collider~(CLIC) and the Cool Copper Collider~(C3). It is planned as a several-stage facility with a few planned upgrades.
The first stage would operate at $250\GeV$ of center-of-mass energy with two interaction regions, allowing for two complementing experiments and a $33.5\km$ tunnel. It would become operational in 2042.
It is designed to collect up to $3$\abinv in the first stage and up to $8$\abinv in the second stage, which will operate at $550\GeV$.
This second stage is necessary to cross-check the different Higgs production modes; at $250\GeV$, the main mode is via the $HZ$ production channel, and at $550\GeV$, the expected dominating mode is $WW$ fusion. Other possibilities involve the measurement of top quark properties at $550\GeV$, with a projected uncertainty of 2\% for the Yukawa coupling when considering $8$\abinv of integrated luminosity. Direct Higgs pair production ($HH$) would also be probed, with a projected uncertainty of 15\% at $550\GeV$ and $4$\abinv.

An additional advantage of the LCF@CERN is the cost that lies well below that of the FCC-ee. The first assessment of the costs for building the LCF@CERN is below $10\,\textrm{bCHF}$.


\subsection{Brazilian scenario}
\label{sec:brsc}
\ifthenelse{\boolean{@draft}}{ \textcolor{red}{responsável: Thiago}\\ }

The Brazilian HEP community is very heterogeneous, albeit well-established, participating not only in collider but also in neutrino and astroparticle physics experiments.
The country participates in all four large LHC collaborations, as well as the ALPHA experiment.
Table~\ref{tab:BrazilAtLHC} shows the statistics of Brazilian participation in the four large LHC experiments; the community is comprised of close to 200 members and 12 institutions.
However, we note that some of the institutional groups are led by a single faculty, so they may have a lower degree of resilience.
\begin{table}[htbp]
       \centering
       \caption{Statistics of Brazilian participation in the four large LHC experiments as of 2024.}\vspace{6pt}
       \begin{tabular}{lcc}
        \toprule
         Experiment   &  Members & Institutions\\
        \midrule
        ATLAS   & 85 & 5\\
        CMS     & 82 & 7\\
        LHCb    & 45 & 3\\
        ALICE   & 24 & 4\\
       \bottomrule
       \end{tabular}
       \label{tab:BrazilAtLHC}
\end{table}
   
In view of Brazil's 2024 accession to CERN as an Associate Member State, it would be natural for the Brazilian community to support the Laboratory's next flagship project.

The FCC option, {\ie} FCC-ee followed by FCC-hh, is a comprehensive plan for the field, which in turn would allow the Brazilian community to unite accelerator and detector R\&D as well as physics goals around the same option. It would, hence, be the favored route.
As discussed in section~\ref{sec:acceleratorRD}, CNPEM has already invested in accelerator R\&D at CERN.
Should LCF@CERN become the viable option for CERN, the Brazilian community would nonetheless be interested in participating in such a facility.

The Brazilian community has both the technical and human capacity to support alternative collider projects such as the LCF@CERN. Brazil can participate across R\&D, construction, operation and analysis. The country offers complementary competencies that enable high-impact contributions on schedules compatible with the post-HL-LHC agenda.

In whichever collider project is ultimately approved, it seems imperative that the Brazilian community aims to participate in as few experimental collaborations as possible, even a single collaboration, to maximize its impact.
    
\subsection{Viability of FCC schedule and constraints on Brazilian contribution}
\ifthenelse{\boolean{@draft}}{ \textcolor{red}{responsável: Thiago}\\ }

We first consider the timeline until the start of construction (2025--2032).
From the point of view of the Brazilian community, the major milestones are those related to the detectors: the submission of the \emph{Expressions of Interest} in 2025 and the preparation of the \emph{Conceptual Design Reports} in 2028--2029.
Unfortunately, the latter period coincides with the LHC Long Shutdown 3: this means that the hardware specialists will have to divide themselves installing the HL-LHC detectors and designing the FCC experiments.
For a small community such as ours, this may prove extremely challenging.
Later steps, such as the preparation of the \emph{Technical Design Reports} and the construction and installation of the detectors should happen during standard HL-LHC data-taking; they would be comparatively easier.

The Brazilian contribution to the FCC detectors may happen on three levels: i) research \& development, ii) construction \& installation, iii) commissioning.
For the first step, the community has extensive experience stemming from its participation in the construction and upgrade of the LHC experiments and the availability of well-furnished experimental facilities in the country.
Brazil already participates in the DRD1 (gaseous detectors) and DRD3 (solid-state detectors) collaborations at CERN. 
However, contributions to the construction step carry the risk of unpredictability of funding and the eventual lack of well-developed industry partners.
The third step may again see significant Brazilian contributions, thanks to the well-trained personnel -- physicists and engineers -- and their availability to relocate to CERN for the FCC startup era.  

With respect to budgetary constraints, there are two main factors to be considered. Firstly, a number of Brazilian funding agencies have formal agreements with CERN, which gives a degree of stability to resource availability;
however, challenges are expected due to the contingently freezing of the funding agencies' budgets.
Second, grants are nominally awarded in local currency (BRL), which has shown to be unstable at times.
For example, the BRL has lost half of its value with respect to the CHF in the 2015--2025 period.


\subsection{Alternative collider options}
\ifthenelse{\boolean{@draft}}{ \textcolor{red}{responsável: Murilo}\\ }

Beyond those discussed in section~\ref{sec:reviewColliderOptions}, alternative collider options include a muon collider, which has been investigated~\cite{Accettura:2023ked}. 
A muon collider holds significant potential for discoveries in the multi-TeV energy range since they provide the full beam energy for interactions. 
For example, a $14\TeV$ muon collider would offer an effective energy reach comparable to a $100\TeV$ FCC.

Another proposed facility is the Circular Electron-Positron Collider~(CEPC)~\cite{CEPCStudyGroup:2023quu}, with a circumference of $100\km$ and a center-of-mass energy of up to $240\GeV$. 
Additionally, plans are to incorporate a Super Proton-Proton Collider~(SPPC) within the same tunnel, capable of reaching a center-of-mass energy of $125\TeV$.

Both of these projects are promising and have their own unique physics goals. 
They are proposed to be constructed before 2040, which overlaps with the timeline of the FCC. 
Given that a CERN-based collider is a priority for the Brazilian experimental HEP community ({\cf} section~\ref{sec:brsc}), 
this timeline conflict makes it less likely for the Brazilian community to participate in the muon collider or CEPC projects in the immediate future.



\section{Non-collider projects}
\ifthenelse{\boolean{@draft}}{ \textcolor{red}{responsável: Gustavo}\\ }

In addition to its strong engagement in collider physics, RENAFAE also supports a broad range of activities in neutrino and astroparticle physics. Brazil plays an active role in the R\&D of liquid argon purification systems for the Long-Baseline Neutrino Facility~(LBNF) and the Deep Underground Neutrino Experiment~(DUNE), with important developments carried out in Brazilian laboratories and partners~\cite{dunebr}. These contributions represent a key milestone in the country’s integration into major global research infrastructure, including related efforts in ProtoDUNE at CERN.
Two experiments have been developed by Brazilian institutes to detect neutrinos from Brazil’s only nuclear power plant complex, the Coherent Neutrino-Nucleus Interaction Experiment~(CONNIE) for the study of low-energy neutrino interactions, BSM searches and reactor monitoring techniques, and the Neutrino-Angra experiment focusing on nuclear monitoring.
Brazilian groups are prominently involved in astroparticle physics, with consolidated participation in the Pierre Auger Observatory in Argentina and growing engagement in future initiatives such as the Cherenkov Telescope Array Observatory~(CTAO).
Moreover, the Brazilian community has been involved in experiments that will take place in the European Spallation Source~(ESS), contributing to their design, simulations and instrumentation.

Beyond fundamental research, CERN has made transformative contributions to medical and industrial technologies\footnote{\url{https://knowledgetransfer.web.cern.ch/applications/healthcare}}. Notably, hadron therapy
offers precise and minimally invasive alternatives to conventional radiotherapy \cite{amaldi2005hadrontherapy}. 
Such technological development and the training of highly qualified personnel are paramount in Brazil.

CERN also advances the production of medical radioisotopes for diagnostic imaging, such as positron emission tomography (PET) and single-photon emission computed tomography (SPECT).
The CERN-MEDICIS project\footnote{\url{https://knowledgetransfer.web.cern.ch/medtech/medicis}} exemplifies the application of accelerator technologies for isotope production, which has enabled breakthroughs in diagnostic imaging. Given Brazil's limited domestic isotope production capacity \cite{mcti2023radioisotopos}, collaboration with CERN-MEDICIS could enhance national capabilities, ensure supply chain strength, and promote local expertise.

CNPEM has demonstrated excellence in magnet technology, as in the construction of the SIRIUS synchrotron light source.
The success of SIRIUS highlights Brazil's ability to develop cutting-edge scientific infrastructure and engage in high-technology industrial innovation. The SIRIUS project also exemplifies a successful partnership model with the private sector, attracting non-public investments in fundamental research and fostering opportunities for technological transfer and economic development. In this context, CNPEM seeks to leverage technological spillovers from ongoing accelerator developments for broader applications, including those in the medical field.

These initiatives reveal the potential of Brazil to contribute meaningfully to multidisciplinary applications of particle physics, and reinforce its role in the global scientific landscape.

\section{Software and Computing}

As part of the Brazilian computational infrastructure, there is a sustained effort to establish a distributed and interoperable Cloud computing network for research, with a framework that interconnects with the \textit{European Grid Infrastructure}~(EGI), promoting collaboration in large-scale modeling and data analysis within the Brazilian community. 

Brazil participates in the Worldwide LHC Computing Grid (WLCG) and has several Tier-2 centers, ensuring large-scale storage and processing resources for LHC experiments for distributed analysis, and interoperability with European and North American sites.


\section{Sustainability}
The Brazilian community recognizes that sustainability should be a key factor when considering future accelerators.
It is engaged in sustainability initiatives as in the development of eco-friendly gases and by establishing energy-efficient data centers.

Alternative gas mixtures are studied, with a focus on Resistive Plate Chambers.
Experimental facilities in Brazil are under construction for the study and development of sustainable gas alternatives, in collaboration with the Gamma Irradiation Facility at CERN~(GIF++)~\cite{gifpp}.
This initiative aims to reduce the environmental impact while advancing detector technology~\cite{Rigoletti:2020b1} and local expertise in Brazil.
It is in line with the guidelines of the European Strategy for Particle Physics and the ECFA Detector R\&D Roadmap~\cite{Group:2021edg}.

Brazil has introduced a new framework to integrate sustainability and artificial intelligence into its digital transformation agenda \cite{MCTI2025,REData2025,MPV1318_2025,PNDataCenters2025}. These initiatives are consolidating a new ecosystem of sustainable scientific computing, anchored in Brazilian national research centers, alongside large-scale facilities like CNPEM. These institutions -- together with Brazil's predominantly renewable energy matrix~\cite{EPE2025} -- position the country as a regional leader in low-carbon, high-performance computing aligned with the global transition towards green digitalization.

\section{Summary}

Brazil’s High-Energy Physics community has expanded its scientific and technological reach through coordinated national efforts led by RENAFAE and INCT CERN-BR, contributing broadly to LHC experiments, detector and accelerator R\&D, computing, and non-collider initiatives in close cooperation with CERN. National infrastructure -- particularly CNPEM -- positions Brazil as a technologically capable partner in accelerator development.

Looking ahead, the community identifies the FCC as the preferred future collider while recognizing that effective participation requires strategic focus and concentration of efforts in a limited number of experimental collaborations to maximize impact. A national governance framework is essential to guide decision milestones, coordinate personnel allocation, and establish phased funding mechanisms that address financial and exchange-rate risks.
Strengthening human capital and technology transfer can be achieved through short-term fellowships for engineers and technologists at CERN, specialized training, and closer collaboration with the Brazilian industry.
These initiatives generate significant socio-economic benefits, including spillovers to healthcare, industry, and local innovation. To further manage financial exposure, flexible contracts or co-financing schemes with adjustable cost shares and payment milestones tied to technical deliverables are recommended. An intensive training plan and prioritization of high-impact contributions would help mitigate limited human capacity, while a personnel-allocation roadmap with rotations and an initial focus on R\&D -- scheduled to avoid HL-LHC installation periods -- would ensure effective use of resources and operational continuity.

By aligning national strengths with CERN’s long-term vision, Brazil reinforces its commitment to a shared scientific future built on collaboration and innovation.

\newpage
\bibliographystyle{JHEP}
\bibliography{refs}

\newpage
\title{Executive Summary after the Open Symposium on the European Strategy for Particle Physics}

\date{November 2025}

\author[a]{(Eds.) U. de Freitas Carneiro da Graça}
\author[b]{G. Gil da Silveira}
\author[c]{C. Jahnke}
\author[d]{A. Lessa}
\author[e]{H. Malbouisson}
\author[f]{E. E. Purcino de Souza}
\author[g]{M. S. Rangel}
\author[h]{T. R. Fernandez Perez Tomei}
\author[a]{A. Vilela Pereira}

\contact[g]{Contact: L. de Paula}

\emailAdd{leandro.de.paula@cern.ch}

\affiliation[a]{
Centro Brasileiro de Pesquisas Físicas~(CBPF)
}
\affiliation[b]{
Universidade Federal do Rio Grande do Sul~(UFRGS)
}
\affiliation[c]{
Universidade Estadual de Campinas~(UNICAMP)
}
\affiliation[d]{
Universidade Federal do ABC~(UFABC)
}
\affiliation[e]{
Universidade do Estado do Rio de Janeiro~(UERJ)
}
\affiliation[f]{
Universidade Federal da Bahia~(UFBA)
}
\affiliation[g]{
Universidade Federal do Rio de Janeiro~(UFRJ)
}
\affiliation[h]{
Universidade Estadual Paulista~(Unesp)
}

\abstract{}

\notoctrue
\maketitle

\pagenumbering{roman}
\setcounter{page}{0}
The Brazilian contribution to the European Strategy for Particle Physics (ESPP) Update was presented to the national community.
Dedicated discussions have been organized since the Open Symposium, during the \textit{National Institute for Science and Technology}~(INCT CERN BR) Workshop in August 2025 and during the Spring Meeting of the Brazilian Physical Society, in October 2025. Possible directions for the future of particle physics in Brazil were discussed.

In addition, a survey was conducted in the community.
%
The results of the survey show that about 85\% of respondents agree that FCC should be supported as the first option.
In the event that it is not viable, roughly 40\% believe that Brazil should support the Linear Collider Facility at CERN (LCF@CERN) or the Compact Linear Collider (CLIC). 
The Muon Collider was highly regarded as an alternative option, receiving support from 25\% of respondents, approximately. Its strong potential for both physics reach and technological innovation was emphasized.
These results support, in view of this committee, the outcome of the previously mentioned organized discussions.

We therefore conclude that the Brazilian HEP community supports the FCC proposal as its main choice for the next major accelerator project at CERN. 
Since becoming an Associate Member State in 2024, the Brazilian community has regarded FCC -- in its both FCC-ee and FCC-hh phases -- as a coherent and forward-looking plan that aligns with national research priorities in accelerator and detector development. 
The project builds on Brazil’s long-standing collaboration with CERN and leverages existing national expertise in superconducting magnet technology, RF systems, and vacuum engineering developed at CNPEM. 
The FCC project is thus seen as Brazil’s preferred path forward, offering a strong foundation for long-term scientific and technological cooperation with CERN.

The LCF@CERN project, which brings together technologies from CLIC and other linear collider initiatives, is viewed by the Brazilian community as a potential alternative should FCC not come to fruition.
Brazilian groups are already contributing efforts to LCF.
While FCC remains Brazil’s primary focus, LCF/CLIC represents a complementary route for advancing precision studies of the Higgs boson and the top quark, making use of mature accelerator technologies and CERN’s existing infrastructure. 
In this scenario, Brazil would remain open to participating in such a facility if it becomes CERN’s next major project.

Finally, the Muon Collider remains a very interesting option due to its technological challenges, which could provide the Brazilian community with a significant role in the international HEP landscape. 
The development of novel ideas and technologies in this context has the potential to open new opportunities for emerging research groups. 
From the physics perspective, the situation is similar, as the Muon Collider offers a vast opportunity for unforeseen advances in both experimental measurements and phenomenological studies.

In summary, the Brazilian HEP community reaffirms its strong support for FCC as the main priority for the next major accelerator project at CERN, recognizing its comprehensive scientific program and its strong alignment with Brazil’s technical expertise and collaborative efforts. 
At the same time, LCF@CERN is viewed as a compelling alternative, particularly given the existing Brazilian contributions to this project.

\end{document}